\documentstyle[12pt,aasms4]{article}

\begin{document}

\title{Fast Time Structure  During Transient Microwave Brightenings: Evidence
 for Nonthermal Processes}
\author{N. Gopalswamy} 
\affil{Department of Physics, The Catholic University of America, Washington DC 
20064}
\author{J. Zhang, M. R. Kundu and E. J. Schmahl}
\affil{Department of Astronomy, University of Maryland, College Park, MD 20742}
\authoremail{gopals@astro.umd.edu}
\begin{center}
and
\end{center}
\author{J. R. Lemen}
\affil{Lockheed Martin Advanced Technology Center, Bldg 252, Department H1-12,
3251 Hanover Street, Palo Alto CA 94304}

\begin{abstract}
Transient microwave brightenings (TMBs) are small-scale 
energy releases from the periphery of sunspot umbrae, 
with a flux density two orders of magnitude smaller than that from a 
typical flare. Gopalswamy et al (1994) first reported
the detection of the TMBs and it was pointed out that the radio 
emission implied a region of very high magnetic field so that
the emission mechanism has to be gyroresonance or nonthermal
gyrosynchrotron, but not free-free emission. It was not possible to decide 
between gyroresonance and gyrosynchrotron processes because of the low time 
resolution (30 s) used in the data analysis. We have since performed a detailed 
analysis of the Very Large Array data with full time 
resolution (3.3 s) at two wavelengths (2 and 3.6 cm) and 
we can now adequately address the question of the emission 
mechanism of the TMBs. We find that nonthermal
processes indeed take place during the TMBs. We present 
evidence for nonthermal emission in the form of temporal 
and spatial structure of the TMBs. The fast time structure
cannot be explained by a thermodynamic cooling time and therefore
requires a nonthermal process. Using the physical parameters obtained from 
X-ray and radio observations, we determine the magnetic field
parameters of the loop and estimate the energy released during the TMBs.
The impulsive components of TMBs imply an energy release rate of $\sim $
$1.3$ $\times$ $10^{22}$ erg s$^{-1}$ so that the thermal energy content
of the TMBs could be less than $\sim$ $10^{24}$ erg.

\end{abstract}

\noindent
Subject headings: Sun: corona --- Sun: flares --- Sun: particle emission --- 
Sun: radio radiation --- Sun: sunspots --- Sun: X-rays, gamma rays

\section{Introduction}

Transient microwave brightenings (TMBs) are small-scale 
energy releases in coronal active regions, first detected by \markcite{gopa1994} 
Gopalswamy et al
(1994) using the Very Large Array\footnote{The Very Large Array
is a facility of the National Radioastronomy Observatory, which is
operated by Associated Universities, Inc., under cooperative agreement 
with the National Science Foundation.} (VLA) at 2 cm wavelength. The TMBs
are compact ($\sim $2 arc sec) sources with duration ranging from less than a 
minute to more than 20 minutes. The typical microwave flux of the TMBs at 
2 cm is nearly two orders of magnitude smaller than that from 
normal flares. The TMBs are also highly polarized, sometimes 
reaching 100\%; they are located close to the spotward footpoints 
of coronal loops connecting the periphery of  
the sunspot umbra to nearby regions of opposite magnetic polarity. 
Gopalswamy et al. (1994, hereafter referred to as Paper 1) interpreted
the TMBs as the radio signatures of small scale heating and/or particle 
acceleration in compact magnetic flux tubes where the magnetic field is 
1200-1800 G. When the microwave observations overlapped with
soft X-ray observations  some TMBs were found to show
X-ray signatures similar to the ones first reported \markcite{shimi1992}
\markcite{shimi1994} by Shimizu et al (1992, 1994). Recently, 
\markcite{shiba1996} Shibasaki (1996) reported
such radio brightenings above a large sunspot, although his events are 
brighter by an order of magnitude than those reported by us.  
In Paper 1, the brightness temperature and polarization of the TMBs were found 
to be consistent with either thermal gyroresonance emission or nonthermal
gyrosynchrotron emission, but not with thermal free-free emission. In this 
letter, we present evidence for both thermal and nonthermal processes during the 
TMBs based on their temporal and spatial evolution.

\section{Data and Results}
 
The TMBs occurred in the vicinity of a large sunspot in AR 7135 (S16W14 at 0 UT 
on 1992 April 24). Observations were made at several wavelengths sequentially.
In Paper 1, a detailed description of the VLA  observations can be found.
We have analyzed these 2 cm observations  with full time resolution (3.3 s) and 
extended the analysis to 3.6 cm wavelength. The resulting radio
images have a spatial resolution of $\sim $$2^{\prime\prime}$ at 2 cm
and $\sim $$3^{\prime\prime}$ at 3.6 cm.
Figure 1 shows the superposition of radio contours on the optical 
image of the sunspot obtained by the SXT aspect sensor on the Yohkoh spacecraft
\markcite{tsun1991} (Tsuneta et al, 1991). The extended emission 
is the gyroresonance emission from the sunspot 
itself. The compact source to the south-east is the TMB. In Fig. 2 we have 
shown another TMB at 2 cm from a different location, to the south-west of 
the sunspot. The 3.6 cm data have confirmed the results of Paper 1 that
the TMBs occur frequently, close to the sunspots. The brightness temperature 
of the TMBs was up to several MK at 3.6 cm and up to 1 MK at 2 cm. In all, 
about two dozen TMBs were observed. Here, we consider only 5 TMBs which form a 
representative sample of all the TMBs.  Table 1 lists the properties of the 
TMBs. 

\subsection{Time Structure of TMBs}

In Fig. 3, we have plotted the intensity at the brightest pixel of each TMB 
listed in Table 1. The time profiles fall into three catagories: impulsive (I) 
(Fig. 3a,b), mixed (I+G) (Fig. 3c,d) and gradual (G) (Fig. 3e) events. 

\noindent
{\bf Gradual Events:} The time profiles of gradual events are similar to the 
gradual rise and fall (GRF) events well known during normal flares, but of 
smaller flux and an overall lifetime exceeding $\sim $ 1 min. Fig. 3e is the 
smooth profile of a gradual event starting around 15:39 UT on 1992 April 24. 
The flux gradually increases by a factor of 3 over a period of several minutes.
The TMB was still in progress when the observation 
ended at 3.6 cm wavelength. When the observation resumed at 3.6 cm an hour 
later, the gradual TMB was gone. 

\noindent
{\bf Impulsive Events:} The impulsive TMBs are very short in overall lifetime, 
typically about a minute. The rise time is of the order of the time resolution 
of the observation. The decay time is relatively longer, but this is probably
due to the presence of a gradual component much weaker than the
impulsive peak. In Fig. 3a,b we have shown two examples of the 
impulsive events: the 22:16 UT event at 2 cm and the 20:12 UT event
at 3.6 cm. The two events occurred at completely different positions 
with respect to the sunspot and had an FWHM of only 10 and 5 s respectively.
The total duration of the TMB was less than 2 min in both cases. 

\noindent
{\bf Mixed Events:} Some TMBs consist of two time scales corresponding to the 
gradual and impulsive components. We refer to these as mixed events. In 
Fig. 3c,d we have presented the time profiles of two TMBs with superposed 
gradual and 
impulsive components. The time structure (with 30 s time resolution) in the 
16:24 UT event was already noted in Paper 1. With 3.3 s time resolution, we 
find that several short time-scale structures are superposed on the relatively 
intense gradual component. This TMB was of sufficiently long duration that it 
continued into the subsequent 3.6 cm observation and showed similar time 
structure (not shown). The impulsive components are seen as modulations on 
the gradual profile. The FWHMs of these impulsive components were again 
$\sim$ 5-10 s. The 18:54 UT event at 3.6 cm is somewhat different in that 
the impulsive components preceded the gradual components as in regular flares.
The TMB consists of three spikes each with a FWHM of $\sim$ 5-10 s, followed 
by a gradual component which lasted for more than 2 min. The gradual component 
itself was superposed by two spikes, each with a duration of $\sim$ 5 s. 

\subsection {Spatial Structure and Polarization}

We compared the spatial structure and polarization of the 
TMB source during impulsive and gradual components. Figure 4 shows the source 
structure of two TMBs during the impulsive and gradual phases. The 20:18 UT TMB
 is impulsive, with an extremely weak gradual component while the 18:53 UT event
has both gradual and impulsive components of comparable brightness (see
Fig. 3a,c for the time profiles of these two events). We see that the radio 
source is rather elongated in one direction during the impulsive component.
We also note that the elongation is in the direction of the magnetic loop 
seen in X-rays as was shown in Paper 1. However, in the decay phase, when the 
impulsive component declined, the source becomes somewhat compact.

Both the impulsive and gradual components are right hand circularly polarized.
Since the sunspot is of positive polarity, both the impulsive and
gradual components have dominant extraordinary mode.
The degree of polarization of the gradual components (shown in parentheses in 
the last column of Table 1) is somewhat larger than that of  the impulsive 
components during any given TMB. This is also true when we compare impulsive and 
gradual events. The difference in polarization between the impulsive and gradual 
components is largest during the 16:24 UT event (88\% for the gradual component 
and 53\% for the impulsive component).

\section{Interpretation}

We interpret the observed time structure as indicative of both thermal
and nonthermal processes during TMBs. The gradual components represent
energy release in the form of heating while the impulsive
components indicate  acceleration of energetic electrons. 
The time scale of the gradual component is consistent with 
typical cooling times of coronal loops. However, fast time structures
cannot be explained by a thermal process because the
cooling time ($\tau_c$) is usually an order of magnitude 
larger than the FWHM of the impulsive components. The cooling time (in seconds) 
in coronal loops is given by \markcite{seri1991} (Serio et al, 1991),
\begin{equation}
\tau_c={120}{L_9}T_7^{-0.5},
\end{equation}
where $L_9$ is the loop half-length in units of $10^9$ cm and $T_7$ is the
temperature of the loop in units of $10^7$ K.  Almost all the TMBs
discussed in this paper came from one footpoint of a single magnetic loop 
structure
which had a typical half-length of $\sim 1.2$ $\times$ $10^9$ cm. The 
temperature of the loop was obtained from soft X-ray observations as $\sim$ 
5 MK. The resulting cooling time is $204$ s which is 40 times larger than the 
FWHM of the impulsive components. Since $\tau_c$ scales
linearly with the loop length for a given temperature, we need  a
loop length much less than an arc sec if the fast time structure were to 
be explained by cooling.

Since the TMBs are located in the vicinity of a large sunspot, the
magnetic field is expected to be very high and is expected to play 
an essential role in the emission process. The relevant thermal emission 
process for the gradual component is gyroresonance emission, since the free-free 
emissionis negligible as shown in Paper 1. The thermal gyroresonance emission 
is also consistent with the high degree of polarization of the gradual 
component.  For the 2 cm emission, the relevant gyroharmonic number is 3, 
corresponding to a magnetic field of 1800 G.  This is also suggested by the 
compact source structure observed for the gradual components. 

For the impulsive components, the relevant emission mechanism is 
optically thin gyrosynchrotron emission from nonthermal electrons. 
In order to explain the fast time structure, the nonthermal
particles must have a lifetime similar to the duration of the
impulsive components. The lifetime of nonthermal particles in a 
coronal loop is determined by the collisional damping time ($\tau_l$):
\begin{equation}
\tau_l = 2 \times 10{^8}{n^{-1}}{E^{3/2}}
\end{equation}
\noindent
where E is the energy (in units of keV) of the nonthermal particles and 
n (in units of cm$^{-3}$) is the thermal electron density in the coronal loop. 
We again make use of the density of the loop ($\sim$ $5 \times$ $10^9$ 
cm$^{-3}$) obtained from soft X-ray observations reported in Paper 1. In order 
to account for the observed duration ($\tau_l$ = $5-10$ s) of the impulsive 
components, we need nonthermal particles in the energy range $10-20$ keV. It is 
significant to note that this is the energy range of nonthermal electrons 
involved in the  production of weak metric type III radio bursts 
(\markcite{lin1981}e.g., Lin et al 1981). If higher energy electrons are produced, all 
of them have to escape from the loop in order to be consistent with the duration 
of the impulsive components.

Gyrosynchrotron emission from such low energy electrons can occur  only at 
the first few harmonics of the gyrofrequency.  At 2 cm, the relevant harmonics 
are  3 to 5 corresponding to a field of 1800 to 1070 G along the loop. We have 
excluded harmonic 2 which would need magnetic fields higher than indicated by 
observations. For 3.6 cm emission, the relevant harmonics are 2 to 4, 
corresponding to field strengths in the range 1500 to 1000 G. The brightness 
temperature contribution at harmonics higher than 5 is 
negligible. Thus one expects nonthermal microwave flux from 
the section of the coronal loop where the magnetic field is 1800 to 1000 G. 
This is why we see the elongation of the source  for the impulsive
components. This is in contrast to the gyroresonance source for which the
(optically thick) emission comes from a single harmonic. For TMBs with impulsive 
and gradual components, the lowest harmonic emission consists of contributions
from both thermal and nonthermal processes. 

Since the spotward leg of the magnetic loop has a high magnetic field and the 
opposite leg connects to a region of very low photospheric field, one expects 
a rapid decline of the magnetic field with distance away from the spot. We can 
determine the magnetic field gradient in the region of 3.6 cm emission as 
$\sim$ 0.1 G km$^{-1}$, corresponding to a magnetic field change from 1500 G 
to 1000 G over a distance of about $6.7^{\prime\prime}$ (see Table 1) along the
loop. This is consistent with the values computed from thermal gyroresonance 
emission of a TMB observed at 2 and 3.6 cm  (Zhang et al 1997).

The 3.6 cm peak brightness temperature ($>$ 10 MK) of the TMB at 20:12 UT 
(see Table 1) is further evidence for nonthermal emission: the electron
temperature of the coronal loop in which the TMB occurred is  only about
5 MK. Unfortunately, there was no X-ray observation at the precise moment of the
TMB although the temperature was measured a few minutes before and after the 
TMB. 

\section{Discussion and Conclusions}

We have presented evidence for nonthermal radio emission during TMBs, in
addition to thermal emission. The nonthermal radio emission is from energetic 
electrons with  energy around 10-20 keV. For both thermal and nonthermal 
emissions reported in this paper, the existence of a strong magnetic field is an 
important factor permitting gyrosynchrotron and thermal gyroresonance emissions 
at low  harmonics. In this respect, these TMBs are unique and are confined to the
neighborhood of large sunspots. The typical 2 cm microwave flux due to 
free-free emission  from the magnetic loop in question is insignificant.
In another study, \markcite{whit1995} White et al. (1995) searched for the 
radio signatures of four X-ray transient brightenings and found that the time 
profiles in X-rays and radio were similar, suggesting plasma heating rather 
than particle acceleration.  It must be pointed out that the events studied by 
White et al. did not originate from the neighborhood of large sunspots.  It may 
be hard to detect nonthermal microwave emission from 10 keV electrons in coronal 
loops away from the sunspot (where the magnetic field is low) due to the following 
reason: microwave flux due to thermal gyroresonance and nonthermal 
gyrosynchrotron flux is insignificant at harmonics above the first few, but the 
low magnetic field means the emission has to be at high gyroharmonics (harmonic 
number 27 is needed for emission at 2 cm in a field of 200 G). Our conclusion is 
consistent with the fact that nonthermal hard X-rays in the energy range 7-10 
keV were detected by CGRO/BATSE \markcite{feff1996} (Feffer, Lin \& Schwartz, 
1996). We predict that BATSE spectroscopy detectors will be able to detect 
nonthermal processes in TMBs close to and away from sunspots while microwave 
instruments can detect only close to the sunspot. Thus lack of nonthermal
radio emission from regions away from the sunspot does not mean that the
energy release is purely thermal. 

Recently, \markcite{gary1997} Gary et al. (1997) studied a larger sample of
TMBs using Owen's Valley Radio Observatory (OVRO) data at many frequencies
in the range 1-18 GHz. Taking advantage of the multifrequency observations, they
determined the spectra of the TMBs and found that some of them did have
nonthermal spectra.  Unfortunately, imaging of the TMBs was not possible using 
the OVRO Solar Array  and we do not know the location of the TMBs with 
respect to the soft X-ray sources. 

The finding that many of the TMBs contain a nonthermal component also
raises an important question whether counting just the thermal
signature such as the soft X-ray brightenings is adequate to 
decide the contribution of these small-scale releases to the coronal
heating. If the released energy goes predominantly into nonthermal
particles which in turn lose their energy to the coronal loop, we may
be undercounting the number of energy input episodes to the coronal loops.
Indeed, it was shown by \markcite{pere1992} Peres et al. (1992) that a coronal 
loop can be  maintained at a steady temperature, provided the time interval 
between two successive energy inputs is less than the cooling time, which is 
typically a few minutes as given by equation (1). 

The typical energy released during a nonthermal pulse can be calculated as 
follows. The volume of the loop determined from X-ray observations is $\sim$ 
2$\times$ $10^{26}$ cm$^3$.  Using the required nonthermal (20 keV) particle 
density of $\sim$ 5$\times$ $10^3$ cm$^{-3}$, we get the total energy released 
during a nonthermal spike as 6.4 $\times$ $10^{22}$ erg, with an energy 
release rate at least 1.3 $\times$ $10^{22}$ erg s$^{-1}$. If all the 
nonthermal electrons lose their energy to the coronal loop, then the thermal 
energy content of the TMBs can be determined from the energy release rate. 
For example, the short duration TMBs lasting for about a minute would carry a 
thermal energy content of $\sim$ 7.8 $\times$ $10^{23}$ erg. This is over 
an order of magnitude smaller than the thermal energy content ($10^{25}$ erg) 
of a typical soft X-ray brightening. On the other hand, the longer duration 
TMBs, such as the 16:24 UT event in Table 1, would have a thermal energy 
content of $\sim$ $10^{25}$ erg. Thus radio observations indicate energy 
releases much lower than what is found from X-ray observations. The TMBs, 
therefore, seem to be good examples of \markcite{park1988} Parker (1988) 
nanoflares and hence may be important in heating at least certain regions of 
the solar corona. It is unclear how the energy is actually released and 
distributed between  particle acceleration and heating during these energy 
release episodes at the smallest scales. This question arises because some of 
the TMBs do not seem to have a gradual component. In order to fully  understand 
this question, we need simultaneous radio and X-ray observations with 
sufficiently high spatial and temporal resolution. 

\acknowledgments

NG and MRK was supported by NASA (NAG-5-6139) and NSF (ATM-901983) grants. 
JRL was supported by NASA contract NAS 8-40801. We thank the anonymous 
referee for suggestions to improve the presentation of this letter.
 
\clearpage
 
\begin{deluxetable}{crrrrrrrrrrr}
\footnotesize
\tablecaption{Properties of Transient Microwave Brightenings}
\tablewidth{0pt}
\tablehead{
\colhead{UT range} & \colhead{$\lambda$} & \colhead{Gradual or} & 
\colhead{$\tau_{spike}$} & \colhead{$\tau_{TMB}$} & \colhead{beam size} & 
\colhead{source size} & \colhead{Flux} & \colhead{$T_b$} & 
\colhead{polarization}
\nl
\colhead{1992 Apr 24} & \colhead{(cm)} & \colhead{Impulsive} 
&\colhead{(sec)} &\colhead{(min)} &\colhead{($arcsec^2$)} 
&\colhead{($arcsec^2$)} 
&\colhead{(SFU)} &\colhead{(MK)} &\colhead{($\%$)}
}
\startdata
$15:39$$-$$15:49$ &$3.6$ &$G$ &$-$ &$>8$ &$3.9$$\times$$2.8$ &$5.3$$\times$
$3.4$ &$0.14$ &$1.3$ &$(62)$ 
\nl
$16:24$$-$$16:44$ &$2.0$ &$I+G$ &$10$ &$14.0$ &$2.0$$\times$$1.5$ 
&$2.4$$\times$$1.8$ &$0.033$ &$0.4$ &$53$$(88)$
\nl
$18:38$$-$$18:56$ &$3.6$ &$I+G$ &$5$ &$3.0$ &$3.0$$\times$$2.5$ &$3.9$$\times$
$2.8$ &$0.15$ &$2.4$ &$46$$(51)$
\nl
$20:12$$-$$20:30$ &$3.6$ &$I+G$ &$5$ &$2.0$ &$3.0$$\times$$2.6$ &$6.7$$\times$
$3.5$ &$1.38$ &$10.2$ &$33$$(53)$
\nl
$22:04$$-$$22:09$ &$2.0$ &$I$ &$10$ &$1.5$ &$2.0$$\times$$1.5$ &$2.8$
$\times$$1.6$ &$0.026$ &$0.3$ &$59$

\enddata
\tablenotetext{}{$\lambda$: Observing  wavelength; $G$: Gradual; $I$: Impulsive;
$\tau_{spike}$: Duration $I$ component; $\tau_{TMB}$: Duration of TMB}
\end{deluxetable}

\clearpage

\figcaption[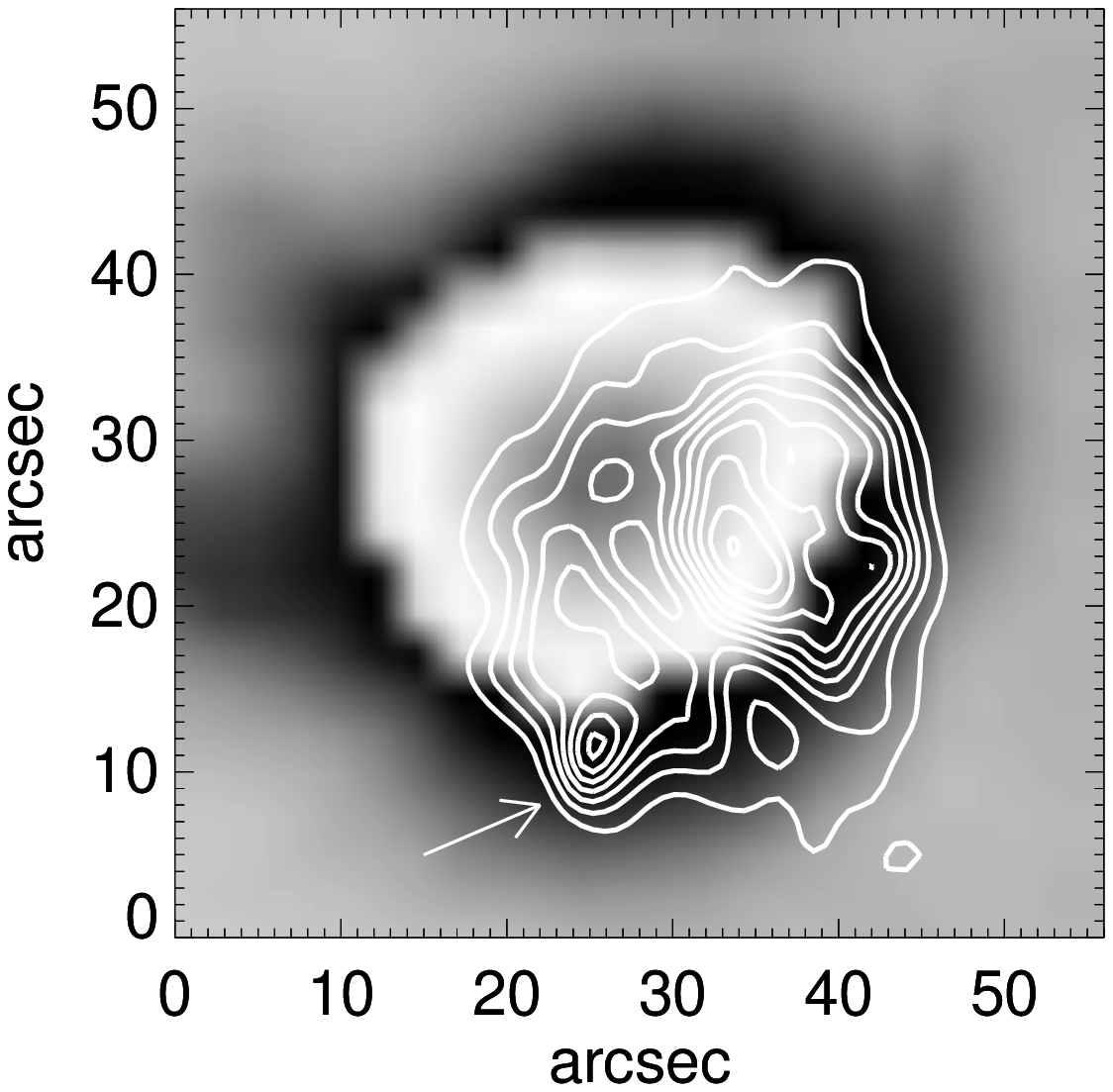]{Overlay of the microwave (VLA 3.6 cm) contours on the 
image of the large sunspot from AR 7135 on 1992 April 24. The sunspot image 
was obtained by the SXT aspect sensor; the bright ring structure near the 
umbra is an artifact because what is displayed is the low 8 bits of the 12-bit 
image. The compact source to the south-east (indicated by the arrow) is the TMB
at 16:34 UT.  North is to the top and east is to the left. The extended
radio contours correspond to the sunspot associated microwave emission 
which is shifted slightly to the south-west of the spot because of its
height in the corona and the angular dependence of gyroresonance emission; the
location of the spot is S16W25. 10, 20, 30, 40, 50, 60, 70, 80, 90, 99\% of the
peak intensity (.0627 sfu/beam); beam size is 
$3^{\prime\prime}$$\times$$2.5^{\prime\prime}$. \label{fig1}}

\figcaption[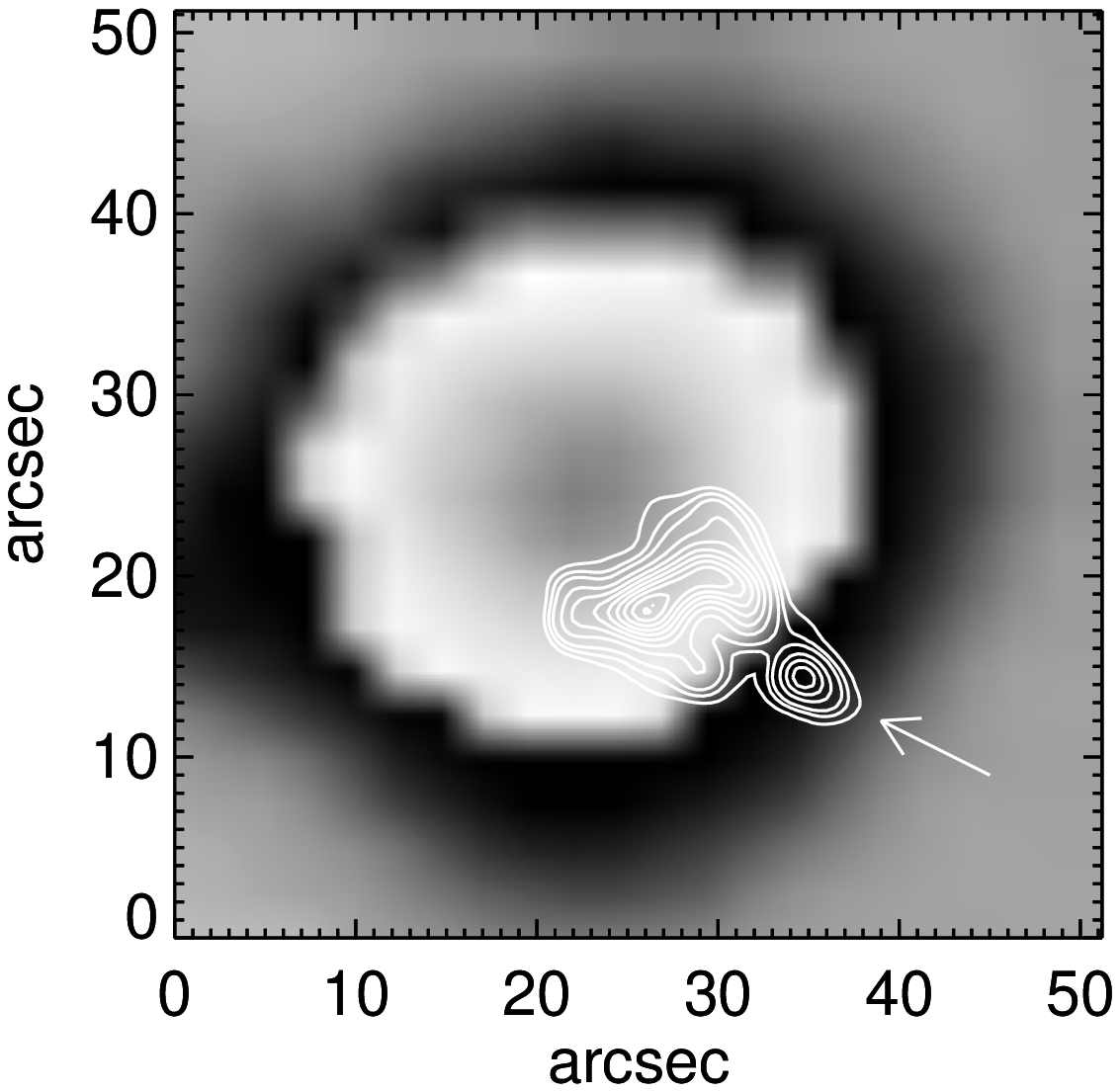]{Overlay of the microwave (2 cm) contours on the 
sunspot image for the 1992 April 24 22:04 UT TMB (indicated by the arrow mark)
which occurred to the south-west of the sunspot. The sunspot is again obtained 
by the SXT aspect sensor with similar artifact as in Fig. 1. North is to the top 
and east 
is to the left. The sunspot associated emission at 2 cm occupies a small 
south-west section of the umbra. The radio contours are at 10, 15, 20,
30, 40, 50, 60, 70, 80, 90, 99\% of the peak intensity (.02496 sfu/beam); beam 
size is $2^{\prime\prime}$$\times$$1.5^{\prime\prime}$.\label{fig2}}

\figcaption[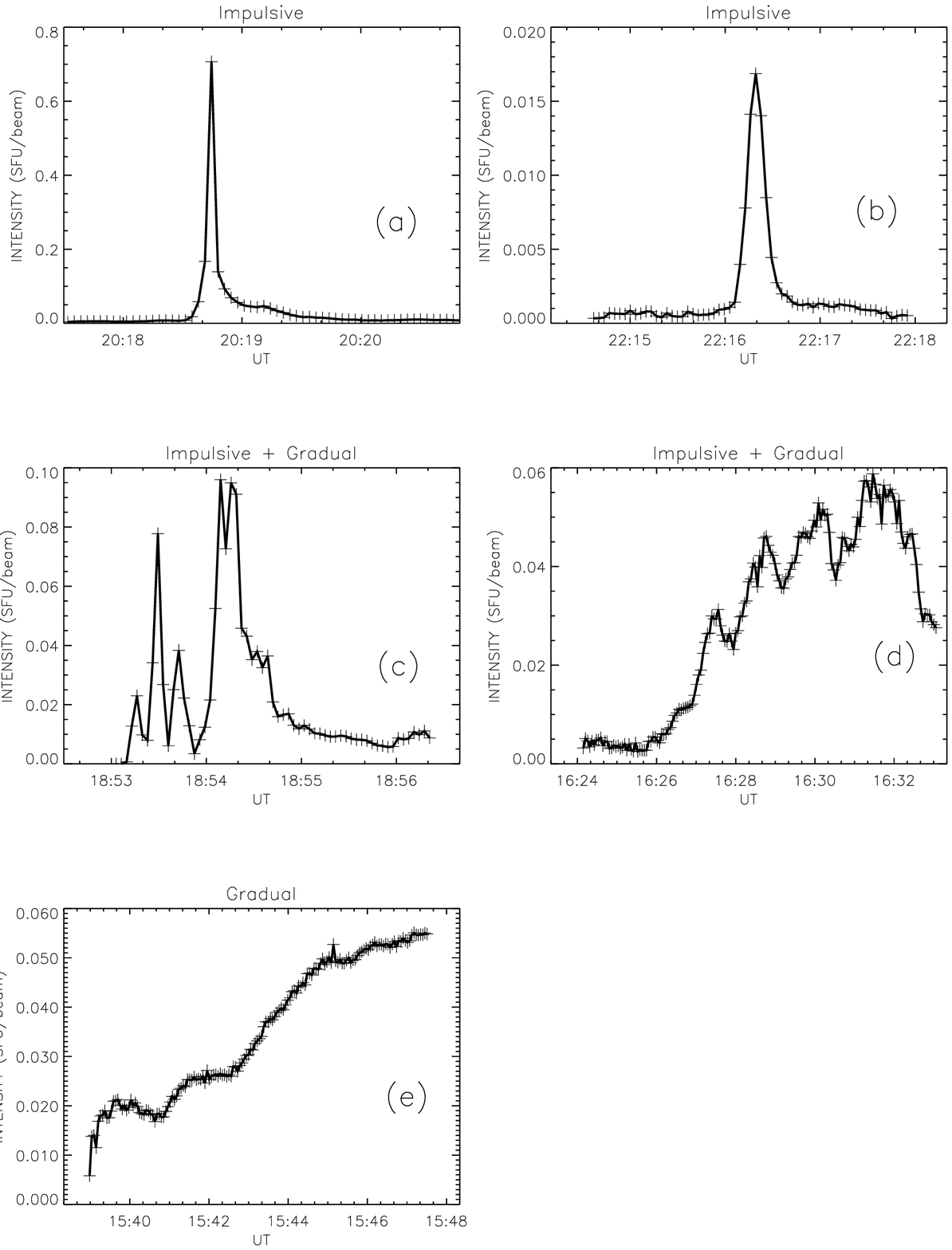]{Microwave time structure of the five TMBs observed on 1992 
April 24. (a) the 22:16 UT TMB at 2 cm and (b) the 20:18 UT TMB at 3.6 cm 
which are purely impulsive. (c) the 18:54 UT TMB at 3.6 cm has 
impulsive components preceding the gradual component.  (d) the 16:24 UT TMB at
2 cm has a dominant gradual component with superposed impulsive components. 
(e) the 15:39 UT TMB at 3.6 cm is a purely gradual event with no significant 
impulsive component. \label{fig3}}

\figcaption[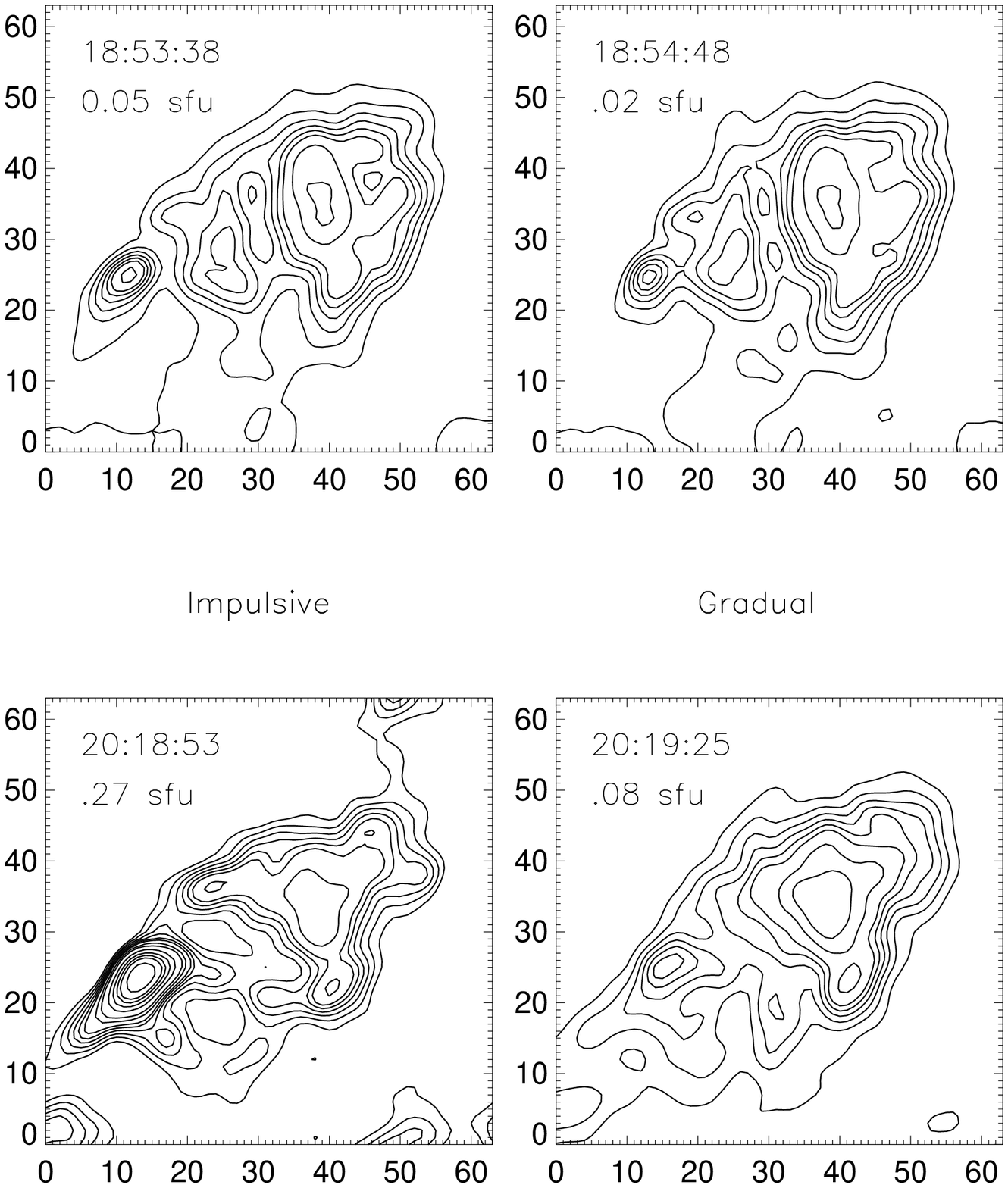]{Radio images of two TMBs observed on 1992 April 24 in 
contour representation. 
The left (right) panel corresponds to the impulsive (gradual) component. Note
that the source is more elongated for the impulsive  component. The peak flux
of each image and the image time are marked on the left hand top corner
of each panel. The contour levels are at 5, 10, 15, 20, 25, 30, 40, 
50, 75, 100, 150, 200 milli sfu/beam. The X and Y axes are in pixel units
 (pixel = 0.6 arcsec at 3.6 cm). \label{fig4}}
 
\end{document}